\documentclass[a4paper,10pt]{article}

\title{\textbf{Indispensable Role of Quantum Theory in the Brain Dynamics}}

\author{Yukinari Kurita\footnote{e-mail:ykurita@phys.ualberta.ca} \\ Department of Physics, P-412, Avadh Bhatia Physics Laboratory,\\ University of Alberta, Edmonton, Alberta, Canada, T6G 2J1 }
\date{ August 24th, 2004}

\begin{document}
\maketitle
\begin{abstract}
Recently, Tegmark pointed out that the superposition of ion states involved in the superposition of firing and resting states of a neuron quickly decohere. It undoubtedly indicates that  neural networks cannot work as quantum computers, or computers taking advantage of coherent states. Does it also mean that the brain can be modeled as a neural network obeying classical physics? Here we show that it does not mean that the brain can be modeled as a neural network obeying classical physics. A brand new perspective in research of neural networks from quantum theoretical aspect is presented.
\end{abstract}

\section{INTRODUCTION}

Although, in most current mainstream biophysics research, the brain is modeled as a neural network obeying classical physics, some researchers have argued that quantum mechanics may play an essential role. On the other hand, recently, Tegmark \cite{tegmark00} pointed out that the superposition of ion states involved in the superposition of firing state and resting state of a neuron quickly decohere. Then he concluded that (a) there is no need to worry about the fact that current neural network simulations do not incorporate effects of quantum coherence, and (b) the only remnant from quantum mechanics is the apparent randomness that we subjectively perceive every time the subject system evolves into a superposition, but this can be simply modeled by including a random number generator in the simulation. The motivation of this paper is to investigate the validity of the conclusion (b).

Over the past few decades, neural networks have provided the dominant framework for understanding how the brain implements the computations necessary for its survival. At the heart of these networks are simplified and static models of nerve cells. However, recently, more and more researchers have been noticing that today's neural network models do not reflect real neurons correctly \cite{koch97, shinomoto98}.

On the other hand, some researchers have argued that quantum theory may play an essential role in the brain dynamics. 

Concerning quantum theory, the past few decades have seen a growing consensus on the decoherence of macroscopic quantum systems \cite{zurek91}. At this point, it was pointed out that environment induced decoherence will rapidly destroy the macroscopic coherent states in the brain \cite{zurek91, zurek03, zeh00, hagan00}. Further, Tegmark \cite{tegmark00} calculated the decoherence rates of the superposition of ion states and found that these states decohere quickly.

The calculation of Tegmark undoubtedly indicates that  neural networks cannot work as quantum computers, or computers taking advantage of coherent states. Does it mean that the brain can be modeled as a neural network obeying classical physics? Here, `classical' means what we observe. This question is still open. It is because research of decoherence cannot define the position of Heisenberg cut, or the observed system whose state vectors seem to collapse by the observation process \cite{zurek91,zeh73,zeh97,zeh96, Saunders02,zurek98} though the research has made great progress not only in theoretical aspect \cite{zurek03,zurek81, zeh70,joos85,joos96, tegmark93} but also in experimental aspect \cite{myatt00}. Hence, decoherence has not defined what classical system is. Nevertheless the following fact has been often missed; decoherence is not the sufficient condition for making quantum system what we observe or classical system, but it is merely the necessary condition. 

If that decoherence occurs in the brain does not mean that the brain can be modeled as a neural network obeying classical physics, it must be possible to suggest brand new perspectives in research of neural networks from quantum theoretical aspect. This is the purpose of the present work.

The rest of the paper is organized as follows. In Section 2, we critically review the work of Tegmark and consider its meaning. In section 3, we see the solution of the question, what defines the position of Heisenberg cut. In section 4, we consider the information processing of the brain. In section 5, we discuss the implications of our results. 

\section{DECOHERENCE IN THE BRAIN}

A quantum superposition of resting and firing states of a neuron involves of the order of a million ions being in a spatial superposition of inside and outside the axon membrane, separated by a distance of the order of $ h \sim 10nm $. Tegmark computed the timescale on which decoherence destroys the spatial superposition states of the ions. The result is about $10^{-20}s$.

The above result undoubtedly indicates neural networks cannot work as quantum computers, or computers taking advantage of coherent states, but it is still an open question whether the brain can be modeled as a neural network obeying classical physics. The reason is decoherence is not the sufficient condition for making quantum system what we observe, or classical system, but it is merely the necessary condition.
 
There is no universally accepted definition of the borderline between a quantum system and a classical system \cite{braginsky87}. In fact, research of decoherence could not have defined the position of Heisenberg cut, or the observed system whose state vectors seem to collapse by the observation process. This problem has been noticed by practitioners of decoherence \cite{zurek91,zeh73,zeh97, zeh96,zurek98}. This means that the relation between what we observe and quantum theory has not been elucidated yet. Nevertheless this fact has been often missed. 

For example, imagine the box which contains gas molecules at 1 atm pressure with a temprature of 300K. In this example, we do not think of delocalized coherent states of each gas molecule exists due to decoherence, but whether each molecule in the box is classical still remains an open question. In other words, whether state vectors of each molecule collapse (or establishes Everett branch) is still an open question. 

Further, the number of ions involved in the firing state of a neuron cannot be rigorously a million, these must be fluctuated. For example, states of the number of ions involved in firing might be the superposition of a million ions and a million plus one ions states. By the way of calculation of Tegmark, we can calculate the decoherence time of the superposition of the states of the number of ions. This time is about $10^{-14}s$. Does this fact mean that the above two states, the state of a million ions involved in firing and the state of a million plus one ions involved in firing, establish different classical worlds? This question is still open.

The above argument indicates that although we have learned decoherence in the brain, it does not necessarily mean that the brain can be modeled as a neural network obeying classical physics. The work of Tegmark has given us only the necessary condition that neural networks are classical. Whether neural networks are classical depends on whether they meat the sufficient condition. 

In next section, we investigate the sufficient condition for making quantum system what we observe, or classical system. This investigation is based on the universality of Schrodinger equation as Everett did \cite{everett57}. This assumption seems to be reasonable in today's situation that we could have made macroscopic coherent states in laboratory \cite{monroe96, arndi99}. 

\section{RELATIVE STATE FORMULATION}

The formulation of observation of quantum theory is the following:

A physical system is described by a state vector $\mid\psi\rangle$,which is an element of a Hilbert space. There are two fundamentally different ways in which the state vector can change.

\begin{description}
\item[Process 1:] The discontinuous change brought about by the observation of a quantity with eigenstates $\mid\phi_1\rangle$,$\mid\phi_2\rangle$,\ldots,in which the state $\mid\psi\rangle$ will be changed to the eigenstate $\mid\phi_i\rangle$ with probability $\mid\langle \phi_i\mid\psi\rangle\mid^2$ .

\item[Process 2:] The continuous, deterministic change of state of an isolated system ,which is described by a state vector $\mid\psi\rangle$ ,with time according to the following equation: $i\frac{\partial \mid\psi\rangle}{\partial t}=H\mid\psi\rangle$
\end{description}

However, Everett \cite{everett57} suggested that there is only the change described by Process 2 in the physical world, and that we do not need Process 1 when we describe an observation process. Further, he insisted that an observed system after observation exists merely as the state relative to an observer(relative state). That is, if the observer described by the state $\mid O\rangle$ observes one of eigenstates [$\mid s\rangle$] of state $\mid\psi\rangle$, after the interaction described by the following equation, the state of the observer who observed certain eigenstate $\mid s\rangle$ is described by $\mid O_s\rangle$. Further, in this observer's world, there is only the relative state $\mid s\rangle$.

\begin{equation}
\label{myeqn1}
\mid\psi\rangle \mid O\rangle=[\sum_{s}c_s\mid s\rangle] \mid O\rangle\\
\to\sum_{s}c_s\mid s\rangle \mid O_s\rangle
\end{equation}

In Everett's paper, we do not find the comment ``the splitting of the world'' \cite{zeh00, vaidman98, tegmark98, lockwood96}. It is an observer that splits. That is, in the case of the above equation (\ref{myeqn1}), the state of an observer $\mid O\rangle$ splits into each observer's state $\mid O_i\rangle$,$\mid O_j\rangle$, \ldots. In other words, the splitting described by the following equation (\ref{myeqn2}) never happens. Here, $\mid R_s\rangle$ is the rest of the world state, which doesn't contain the state of the observer $\mid O_s\rangle$ and the state of the observed system $\mid s \rangle$.

\begin{equation}
\label{myeqn2}
\sum_{s}c_s\mid s\rangle \mid O_s\rangle \mid R_s\rangle
\end{equation}

There are some problems, however. Now we explain this by an model that position states of a system are observed; here both the observer and the system are assumed to be stable. Further, the observer is assumed to be described by a macroscopic localized state $\mid \Psi \rangle$, and the position states of the system are assumed to be described by a superposition of macroscopic localized states $\mid \psi_- \rangle $ and $\mid \psi_+ \rangle $, which are spatially separated from each other. As the observation process, for simplicity, the observer is assumed to effectively interact with the position states of the system through photons. To be more precise, photons effectively interact with the position states of the system, and then the photons effectively interact with the observer.

Here we define effective interaction between matter and photons. The condition of effective interaction is the following; Here, $ \mid \chi \rangle$ is the initial state of the photons and S is scattering matrix of the matter and the photons.
\begin{equation}
\mid \langle \chi \mid S \mid \chi \rangle \mid^2 \sim 0
\end{equation}

At a first step, let us consider that the observer effectively interacts with the position state $\mid \psi_- \rangle $. This condition is the following:

\begin{eqnarray}
\mid \langle \chi \mid S_s \mid \chi \rangle \mid^2 \sim 0\\
\mid \langle \chi (iT_s^{\dagger})\mid S_o \mid (-iT_s)\chi \rangle \mid^2 \sim 0\\
S_s \equiv 1-iT_s
\end{eqnarray}

In the above, the wavelength of photons $\lambda$ is much shorter than the system. $\mid S_o \rangle $ is the scattering matrix of the observer and the photons. $\mid S_s \rangle$ is the scattering matrix whose inputs are momentum state of the photons and each $\lambda^2 cm^2$ of surface area state of the system. 
 
The following could describe the interaction process. Here observer's state $\mid \Psi_- \rangle $ correlates with the position state $\mid \psi_- \rangle $.
\begin{eqnarray}
\label{myeqn3}
\mid \psi_- \rangle \mid \Psi \rangle \to \mid \psi_- \rangle \mid \Psi_- \rangle 
\end{eqnarray}

Similarly, the following could describe the process which the observer effectively interacts with the position state $\mid \psi_+ \rangle $. Here observer's state $\mid \Psi_+ \rangle $ correlates with the position state $\mid \psi_+ \rangle $.

\begin{eqnarray}
\label{myeqn4}
\mid \psi_+  \rangle \mid \Psi \rangle \to \mid \psi_+ \rangle \mid \Psi_+ \rangle 
\end{eqnarray}

From the above two processes, we can guess the process that the observer effectively interacts with the superposition of position states of the system. Roughly speaking, it could be expressed by adding the above two equations(\ref{myeqn3}, \ref{myeqn4}). ( Here, we neglect the interaction between state $\mid \psi_- \rangle $ and state $\mid \psi_+ \rangle $ through photons.)

\begin{eqnarray}
\label{myeqn5}
(\mid \psi_- \rangle +\mid \psi_+ \rangle) \mid \Psi \rangle \to \mid \psi_- \rangle \mid \Psi_- \rangle +\mid \psi_+ \rangle \mid \Psi_+ \rangle 
\end{eqnarray}

In the above equation, position states of the system $\mid \psi_- \rangle $ and $\mid \psi_+ \rangle$ are defined as states relative to the observer's state $\mid \Psi_- \rangle $ and $\mid \Psi_+ \rangle $, respectively provided that the observer's states are mutually orthogonal. However this is not the whole story. We can expand the state of the system ($\mid \psi_- \rangle +\mid \psi_+ \rangle $) as follows. 

\begin{eqnarray}
\label{myeqn6}
\mid \psi_- \rangle +\mid \psi_+ \rangle =\frac{\mid A+\rangle - \mid A-\rangle}{\sqrt{2}}+\frac{\mid A+\rangle + \mid A-\rangle}{\sqrt{2}}
\end{eqnarray}

Where $\mid A+\rangle $ and $\mid A-\rangle $ are different eigenstates of the system. In this case, the above equation(\ref{myeqn5}) could be described by the following. 

\begin{eqnarray}
\mid \psi_- \rangle \mid \Psi_- \rangle +\mid \psi_+ \rangle \mid \Psi_+ \rangle \\
=\frac{\mid A+\rangle - \mid A-\rangle}{\sqrt{2}}\mid \Psi_- \rangle +\frac{\mid A+\rangle + \mid A-\rangle}{\sqrt{2}} \mid \Psi_+ \rangle\\
\label{myeqn11}
=\mid A-\rangle \frac {\mid \psi_+ \rangle-\mid \psi_- \rangle }{\sqrt{2}}+ \mid A+\rangle \frac {\mid \psi_+ \rangle+\mid \psi_- \rangle }{\sqrt{2}}
\end{eqnarray}

In the above equation, states of the system are defined as states relative to the observer's states $\frac{\mid \psi_+ \rangle-\mid \psi_- \rangle}{\sqrt{2}}, \frac{\mid \psi_+ \rangle+\mid \psi_- \rangle}{\sqrt{2}}$. Therefore, it is not clear what state is observed. This is, what is called, preferred basis problem \cite{Saunders02,zurek81,lockwood96,Wallace02,Wallace03}.

The above discussion, however, misses decoherence of the system. Macroscopic superposition of different spatially localized states decohere by coupling with its environment \cite{joos85, joos96}. Hence the expansion (\ref{myeqn11}) is impossible. Let us assume that the environment is gas molecules. In this case, the interaction process is thought to be similar to the process of the equation (\ref{myeqn5}) even though the composite system of the observer, the system, and the photons is not an isolated system. It is because the interaction between the photons and the gas is negligible. Hence the interaction between the observer and the system is the following. 

\begin{eqnarray}
(\mid \psi_- \rangle \mid \Xi_- \rangle +\mid \psi_+ \rangle \mid \Xi_+ \rangle) \mid \Psi \rangle \\
\label{myeqn7}
\to \mid \psi_- \rangle \mid \Xi_- \rangle \mid \Psi_- \rangle +\mid \psi_+ \rangle \mid \Xi_+ \rangle \mid \Psi_+ \rangle \\
\langle \Xi_- \mid \Xi_+ \rangle \sim 0
\end{eqnarray}

$\mid \Xi_- \rangle$ and $\mid \Xi_+ \rangle $ are states of the environment of the system. Before the interaction, states of the environment $\mid \Xi_- \rangle $ and $\mid \Xi_+ \rangle $ correlate with the position states of the system $\mid \psi_- \rangle $ and $\mid \psi_+ \rangle $, respectively. After the interaction, states of the observed system are well defined as states relative to the observer's states provided that the states of the observer are mutually orthogonal. It is because the interaction between the observer and each microscopic states of the gas molecules which caused the decoherence is not effective, in other words, the observer selectively interacts with position states of the system. After the interaction, the observer cannot expand the composite system of the system and its environment in different way any more. 

Hence, when an observer effectively interacts with certain states of a system whose superposition was decohered by coupling with its environment, the observer splits into states each of which is relative to one of the decohered states of the system. 

The above is not the only condition that the observer splits. If the observer A effectively interacted with a certain state of the system B, the observer A splits when the state of the system B splits. In the following, $\mid \Psi_A \rangle $ and $\mid \Psi_B \rangle $ are the state of the observer A and the state of the system B respectively. $\mid \Psi_B^+ \rangle$ and $\mid \Psi_B^- \rangle$ are the state of the observer A and the state of the system B after splitting. $c_1$ and $c_2$ are normalization factors of $\mid \Psi_B^+ \rangle$ and $\mid \Psi_B^- \rangle$
 
\begin{eqnarray}
\mid \Psi_A \rangle \mid \Psi_B \rangle \to c_1 \mid \Psi_A \rangle \mid \Psi_B^+ \rangle+c_2\mid \Psi_A \rangle \mid \Psi_B^- \rangle 
\end{eqnarray}

From this point of view, we conclude the following; 

\begin{description}

\item[Condition of Split;] (a)An observer splits into relative states to certain states of an system when the observer selectively interacts with the certain states whose superposition was decohered by coupling with its environment. (b)An observer splits into states relative to certain states of an system whose state effectively interacted with the observer when the system splits into the certain states. 

\end{description}

In the above Condition of Split, the observer is any matter. In other words, we can regard any matter as the observer. It is because Condition of Split does not necessarily reflect observation process, but it is universally useful to define the relative states to the system. 

Hence even if the split is caused by the above condition (a), this is not necessarily the phenomena which is followed by an observation. In an observation, an observer splits into distinguishablly different states each of which corresponds to one of decohered states of an observed system. For example, in the above equation (\ref{myeqn7}), $\mid \Psi_- \rangle $ and $\mid \Psi_+ \rangle $ are distinguishable each other, or mutually orthogonal. If this condition is not meant, the states of the system relative to the observer's states cannot be defined as mutually distinguishable states, and to the observer, different states of the system still coexist. 

Let us assume the observer is human being in the above model. In this case, the skin of the observer effectively interacts with the gas molecules and the observer feel the pressure of the gas. The skin also could effectively interact with the position states of the system through a part of the gas molecules. Hence the state of the skin could split into relative states to the position states of the system. However these relative states are not mutually distinguishable, hence the position states of the system relative to the skin cannot be defined as mutually distinguishable states. One the other hand, the above interaction through the photons, the states of the retina of the observer splits into relative states to the position states of the system, and these relative states are mutually orthogonal. Hence, the position states of the system relative to the retina can be defined as mutually distinguishable states.

The following is the definition of observation in which certain state vectors of an observed system seems to collapse.

\begin{description}

\item[Condition of Observation;] observation in which certain state vectors of an observed system seem to collapse is selective interaction between an observer and the certain states which are cut from states of the environment of the system and is the interaction by which the state of the observer splits into states which are mutually orthogonal and each of which correlates with each of the certain state vectors.

\end{description}

Here, in an observation, the orthogonality of the states of the environment of an observer is not necessary. Because of the success of explaining preferred basis problem by environment induced decoherence, the orthogonarity of the states of the environment of an observer or the brain has been often assumed \cite {tegmark00, zurek91, lockwood96}. In the above equation (\ref{myeqn7}), however, the states of the system are well defined as relative states to the observer, and the states of the observer are well defined as relative states to the system without the condition of orthogonality of the state of the observer's environment. 
 
Using the above definition of observation, we can define the position of Heisenberg cut, or the observed system whose state vectors seem to collapse by the observation process. The observed system is the quantum system each of whose decohered states is relative to each of mutually orthogonal states of an observer. This is Copernian change in research of decoherence. Here we never need to define, a prior, an observed system and its environment. An observed system and its environment are defined by the interaction with an observer. In the above case, except for the position states of the system, there are no states which effectively interact with the observer and which correlates with each of the mutually orthogonal states of the observer.

One problem still remains; how can we define observed systems and an observer in the brain? We cannot use an external observer to define observed systems in the brain because the brain not only processes the information but also interprets the pattern of the activity \cite{pribram91}. Next section, we consider this problem.

\section{DYNAMICS IN THE BRAIN}

According to paradigm of today's brain science, a perception depends on the simultaneous, cooperative activity of neurons \cite{freeman91, edelman00}. Hence we cannot regard certain object in the brain as an observer in the brain, but  dynamics of the brain must compose an observer in the brain. This is an important point. As long as we use an external observer, we can define a classical system in the brain easily. A classical system is the system whose states is relative to an external observer. Without an external observer, however, we cannot define an observed system and its environment in the brain, and hence we cannot define a classical system. 

In the following, firstly, we think about the process of signal transmission in the brain. To do so, we think about electrical synaptic transmission and chemical synaptic transmission because there are two types of neuron in the brain, which take advantage of these transmissions \cite{nicholls00}. Then, we consider the information processing of the brain. 

In the following argument, at least a part of neurons in the brain is assumed to fire quantum probabilistically. 

\subsubsection*{electrical synaptic transmission}

Although it was considered, a priori, that electrical synaptic transmission would not be present in mammalian forms, it has been found to play an important role in the information processing of the brain \cite{llinas85, bennett04}. In this transmission, current flows directly from one neuron to another through connexons, intercellular channels that cluster to form gap junctions \cite{nicholls00}.

Now let us think the following example; presynaptic neuron A fires quantum probabilistically, and the neuron A is electrically connected to postsynaptic neuron B. Further, for simplicity, the neuron B fires by the electrical signal from the neuron A.

In the above case, when the neuron A fires quantum probabilistically, the superposition of ion states involved in the superposition of firing state and resting state of the neuron A decohere because of the result of Tegmark. Then, through electrical synaptic transmission, the neuron B splits into 2 states because of the above Condition of Observation \footnote{Strictly speaking, there is a possibility that the neuron B splits into relative states to different number of ions involved in firing of the neuron A. However these relative states are not mutually distinguishable.}. These 2 states are the states; the state which correlates with the ions involved in firing state of the neuron A and the state which correlates with the ions involved in resting state of the neuron A. Hence the superposition states of the neuron B are generated, one of which is firing state of the neuron B and the other of which is resting state of the neuron B. Then the superposition of ion states involved in the superposition of firing state and resting state of the neuron B decohere quickly. 

Firing state of the neuron A and the neuron B, however, goes back to resting state quickly. According to the paradigm of today's brain science, a neural network takes advantage of synaptic efficacy to memorize an event, and it needs repetitive activity to produce long term synaptic efficacy \cite{nicholls00}. This indicates that a neural network cannot keep memory of one firing event for a long time. Hence after firing state of a neuron goes back to resting state, whether it fired might not influence on function of a neural network.

However there is a possibility that whether it fired could influence on function of a neural network. This is the case that the membrane of a neuron which experienced firing and that of the neuron which did not experience firing are separable because the separability means a signal path is divided into 2. This separability is meant if a neuron itself or its membrane can keep the memory of the firing event after it goes back to resting state. Now we consider this possibility. Let us think about the following process. 

\begin{eqnarray}
\label{myeqn8}
\mid \mathrm{neuron}_{resting}\rangle \mid E\rangle \\
\label{myeqn9}
\to \mid \mathrm{neuron}_{firing} \rangle \mid E_f\rangle+\mid \mathrm{neuron}_{resting} \rangle \mid E_r \rangle \\
\label{myeqn10}
\to \mid \mathrm{neuron}_{resting2} \rangle \mid E_{r2}\rangle+\mid \mathrm{neuron}_{resting} \rangle \mid E_r\rangle 
\end{eqnarray}

where $\mid \mathrm{neuron}_{firing}\rangle$ and $\mid \mathrm{neuron}_{resting} \rangle $ are neuron's (or its membrane's) state in firing and neuron's (or its membrane's)state in resting, respectively. Further, $\mid \mathrm{neuron}_{resting2}\rangle$ is the neuron's state in resting which has experienced firing, or neuron's state which the neuron's state in firing goes back to. $\mid E\rangle$, $\mid E_f\rangle$, $\mid E_r\rangle$, and $\mid E_{f2}\rangle$ are the states of the environment of a neuron, each of which is composed of states of inter neural space and outer neural space. Here, neuron's state in firing is any macroscopic state of neuron's membrane when it fires, and the state contains many different microscopic states of each atom, such as position, energy, spin. Similarly, neuron's state in resting is any macroscopic state of neuron's membrane when it is resting, and the state contains many different microscopic states of each atom, such as position, energy, spin. 

Here, if the condition $\langle \mathrm{neuron}_{resting2}\mid \mathrm{neuron}_{resting}\rangle \sim 1 $ is meant, the membrane of a neuron which experienced firing and that of the neuron which did not experience firing are not separable. \footnote{Strictly speaking, the above expressions (\ref{myeqn8}), (\ref{myeqn9}), (\ref{myeqn10}) are not correct. The above neuron's states are never pure states because they are macroscopic and their microscopic states strongly correlate with their environment. However, we adapt the above expression for simplicity. Further, $\langle \mathrm{neuron}_{resting2}\mid \mathrm{neuron}_{resting}\rangle \sim 1 $ means the sum of inner product of each pure state.}. This is a reasonable condition because of the stability of a neuron; 

A neuron strongly interacts with its environment constantly and neuron's state in resting and neuron's state in firing are composed of many microscopic states. Hence the microscopic states continue to change by the interaction with its environment. The content of the environment is mainly water molecules because concentration of ion is much less than $10^{-3}$ \cite{katz66}. Further, the entrances of ion channels, which mediate the flow of ions, cover only about $10^{-4}$ of the membrane area \cite{hoppe83}. Here, the difference between the state of the environment involved in firing and the state of the environment involved in resting can be regarded as the difference of ion states because the state of the water molecules which strongly interact with a neuron is hardly changed by firing. The microscopic change of neuron's state by the interaction with ions is negligibly small compared to its constant change by the interaction with water molecules. Hence, the difference between neuron's interaction with the ions involved in firing and resting is absorbed into the fluctuation of neuron's microscopic states by the interaction with water molecules.

Hence, after going back to resting, the membrane of a neuron which experienced firing and that of the neuron which did not experience firing are not separable. This means that after firing state of a neuron goes back to resting state, the neuron has single signal path because the neuron's state in resting cannot be divided into 2 states, one of which experienced firing and the other of which did not experience firing. Hence after firing state of a neuron goes back to resting state, whether the neuron fired is almost irrelevant to function of a neural network.

As we mentioned, a perception depends on the simultaneous, cooperative activity of neurons. Further, majority of neuroscientists believe the detailed pattern of spikes is largely irrelevant \cite{koch97, softky93}. Therefore, decohered superposition of ion states involved in firing state and resting state coexist in the brain or in a same perception. Nevertheless an external observer observe one of the 2 states in measurement of spike in the brain because the measurement meets Condition of Observation. 

\subsubsection*{chemical synaptic transmission}

In this transmission, depolarization of the presynaptic nerve terminal triggers the release of chemical transmitters, which interact with receptors on the postsynaptic neuron, causing excitation or inhibition \cite{nicholls00}. The release process takes advantage of exocytosis; vesicles packed chemical transmitters dock in synapse and some of vesicles release chemical transmitters. Although the detail of recycling of vesicles is still debated \cite{henkel96, henkel95, gandhi03, aravanis03}, there is the possibility that the state of the vesicle which has experienced exocytosis and the state of the vesicle which has not experienced exocytosis are delocalized by disconnected from synapse. In this case, the 2 states of the vesicle quickly decohere because a vesicle is a macroscopic object. 

Now let us think the following example; presynaptic neuron A fires quantum probabilistically, and the neuron A is connected to postsynaptic neuron B. Further, for simplicity, the neuron A release chemical transmitters when it fires, and the neuron B fires by receiving chemical transmitters from the neuron A.

After the neuron B fires, by Condition of splits (a) or (b), the neuron A and the neuron B permanently split into 2 states, one is the state which is relative to the state of the vesicle which has experienced exocytosis and the other is the state which is relative to the state of the vesicle which has not experienced exocytosis. However, it does not necessarily mean the brain splits into the 2 states, or 2 different perceptions happen.

As we mentioned, firing state of a neuron goes back to resting state quickly, and a neural network takes advantage of synaptic efficacy to memorize an event. In other words, memory is not composed of the coordinate of vesicles, but it is composed of synaptic efficacy. Further, as we mentioned, after going back to resting, the membrane of a neuron which experienced firing and that of the neuron which did not experience firing are not separable. Hence after firing state of a neuron goes back to resting state, whether the neuron fired is almost irrelevant to function of a neural network.

Now is the time that we confirm the meaning of relative state. As we mentioned, the neuron A and the neuron B permanently split into 2 states, but these 2 states are not mutually orthogonal. Hence, delocalized states of the vesicle relative to the neuron A or neuron B cannot be defined as mutually distinguishable states. The split defines merely relative states to the vesicle. However this relative relation is irrelevant to signal transmission. Signal transmission depends on whether a neural network memorizes firing events. It also depends on whether the above 2 states of the membrane of a neuron are separable because if they are separable, two different paths of signals are generated.

From the above argument, we conclude that the states of the delocalized vesicle coexists in the brain or in a same perception. Nevertheless an external observer observe localized state of the vesicle because the observation meets Condition of Observation \footnote{This does not mean that we can observe the superposition of delocalized vesicle if the delocalized states of the vesicle do not decohere. It is because states of the delocalized vesicle interact with different positions of retina through photons, and the states of the different positions of retina decohere quickly. This situation is the same as that of experiment of a diffraction grating. In this experiment, states of photons which interact with different positions of retina decohere quickly.}.

\subsubsection*{information processing of the brain}

In the above, we have seen that firing state and resting state coexist in the brain. This leads to the hypothesis; the brain takes advantage of decohered superposition of ion states involved in firing and resting. At first glance, this hypothesis seems to deny 2 main doctrines on the neural code by which information is transferred through the cortex; statistical average of spike code and precise spike time code. 

The above hypothesis, however, could reconcile the fact that the 2 main doctrines are consistent with experiments \cite{koch97,softky95, shadlen94}. Even though the brain takes advantage of the decohered superposition, each state of the superposition is followed by quantum probability, and this probability is related with the observation of statistic quantity of spikes. Hence, the experimental results which are consistent with the doctrine of the statistical average could be consistent with the hypothesis, too. Further, since the brain takes advantage of the decohered superposition states, it can take advantage of timing of firing. Hence, the experimental results which are consistent with the doctrine of the precise spike time code could be consistent with the hypothesis, too.

Here, it must be emphasized that the phenomena which compose brain activity are different from phenomena in the brain which are observed by an external observer. 

The argument, an external observer should be avoided, has been argued in the classical context, too \cite{edelman00}. According to Edelman, differences between activity patterns in the brain should be assessed only with reference to the system itself. To do so, he suggested to divide the system in two and to consider how one part of the system affects the rest of the system, and vice versa.

The above suggestion is useful in quantum context, too. Although the brain takes advantage of decohered superposition states, the probability of firing must depend on synaptic connection, and synaptic connection reflects one neuron's dependency on another. Hence, to analyze the brain dynamics from inside of the brain, we need the tools which estimate one system's dependency on another, such as mutual information \cite{edelman00, zurek00}. 

\section{DISCUSSION}

So far, we have not considered whether the final output of the brain is only one. This is an open question. There is a possibility that the final output is not only one. If exclusive states of cooperative activities of neurons produce different perceptions, the brain or a perception may split. In addition, if exclusive states of cooperative activities of neurons cause to produce different long term memories, the brain or a perception may split. This could be useful for explaining the free will.

In the above, we have seen that the phenomena which compose brain activity are different from phenomena in the brain which are observed by an external observer. It is important for not only brain science but also foundation of physics. Apart from well devised experiment \cite{plaga97}, it has been insisted that different `interpretation of quantum theory' cannot be distinguished because of decoherence \cite {joos96, tegmark93}. This is the reason why majority of physicists do not regard observation problem of quantum theory as science. The brain activity, however, is composed of the phenomena which have not been observed in outside world. Hence, foundation of physics has no more been the subject of philosophy of science. It is subject of both brain science and physics though it is still a difficult problem to find experimental proof of the existence of the phenomena.

In the above argument, we assumed the following two. Firstly, we assumed that at least a part of neurons in the brain fire quantum probabilistically. There is no proof of this assumption though there are some theoretical tentativeness \cite{donald90, beck92, roy04}. Secondly, we assumed the conditions $\langle \mathrm{neuron}_{resting2}\mid \mathrm{neuron}_{resting}\rangle \sim 1 $ without rigorous quantitative estimation. There is a possibility that these assumptions are falsified. Even if the assumptions are falsified, however, this does not necessarily mean that firing state and resting state compose different perceptions, but there is a possibility that firing state and resting state coexists in a same perception. Anyway, there is little doubt that the following fact is crucially important for brain research; the phenomena which compose brain activity are different from phenomena in the brain which are observed by an external observer.

Further, even if the above hypothesis is falsified, the hypothesis suggests the possibility of a brand new computer, which is different from both classical computers and quantum computers, which take advantage of coherent states. This new computer takes advantage of decohered superposition.

\subsubsection*{Acknowledgements}
The author would like to thank D.\ N.\ Page for discussion, and F.\ C.\ Khanna, J.\ A.\ Tuszynski, M.\ J.\ Donald, M.\ Tegmark, S.\ Shinomoto and T.\ Kaneko for useful comments. The author also would like to thank Department of Physics in University of Alberta for hospitality. 

\addcontentsline{toc}{section}{References}
\bibliographystyle{abbrv}

\begin{thebibliography}{19}
\bibitem{tegmark00}M.\ Tegmark, ``Quantum computation in Brain Microtubules? Decoherence and Biological Feasibility '', \textit{Phys.\ Rev.E.}, \textbf{ 61,} 4194 (2000).

\bibitem{koch97}C.\ Koch, ``Computation and the single neuron '', \textit{Nature(London)}, \textbf{385,} 207 (1997).

\bibitem{shinomoto98}S.\ Shinomoto, ``What should we investigate in neuroscience?'', \textit{Sinkeikairogakkaishi}, \textbf{Vol.5 No.2,}(1998).

\bibitem{zurek91} W.\ H.\ Zurek, ``Decoherence and the transition from quantum to classical'', \textit{Phys.\ Today}, \textbf{ 44,} 36 (1991).

\bibitem{zurek03} W.\ H.\ Zurek, ``Decoherence, einselection, and the quantum origin of the classical'', \textit{ Rev.\ Mod.\ Phys. }, \textbf{ 75,} 715 (2003).

\bibitem{zeh00}H.\ D.\ Zeh, ``The problem of conscious observation in quantum mechanical description'', 2000, quant-ph/9908084.

\bibitem{hagan00}S.\ Hagan, S.\ R.\ Hameroff, J.\ A.\ Tuszynski, ``Quantum computation in brain microtubles? Decoherence and biological feasibility'',quant-ph/0005025.

\bibitem{zeh73} H.\ D.\ Zeh, ``Toward a Quantum Theory of Observation'', \textit{Found.Phys.}\textbf{Vol.3,}109(1973).

\bibitem{zeh97}H.\ D.\ Zeh, ``What is achieved by decoherence '', \textit{New Development on Fundamental Problems in Quantum Physics}(Kluwer Academic, 1997).

\bibitem{zeh96} H.\ D.\ Zeh, ``The program of decoherence: Idea and Concepts'',\textit{Decoherence and the Appearance of a Classical World in Quantum Theory}(Springer, 1996).

\bibitem{Saunders02}S.\ Saunders, ``Time, Quantum Mechanics, and Probability", \textit{Synthese}\textbf{235,}19(1995). 

\bibitem{zurek98} W.\ H.\ Zurek, ``Decoherence, einselection, and the existential interpretation (the rough guide)'', \textit{Phil.\ Trans.\ R.\ Soc.\ Lond.\ A. }, \textbf{ 356,} 1793 (1998).

\bibitem{zurek81} W.\ H.\ Zurek, ``Pointer basis of quantum apparatus: Into what mixture does the wave function collapse?'', \textit{Phys. Rev.}\textbf{ D24,}1516(1981).

\bibitem{zeh70} H.\ D.\ Zeh, ``On the interpretation of measurement in quantum theory'', \textit{Found.Phys.}\textbf{Vol.1,}69(1970).

\bibitem{joos85} E.\ Joos, and H.\ D.\ Zeh, ``The emergence of classical properties through interaction with the environment'', \textit{Z.Phys.}\textbf{ B59,}223(1985).

\bibitem{joos96} E.\ Joos, ``Decoherence through interaction with the environment'', \textit{Decoherence and the Appearance of a Classical World in Quantum Theory}(Springer, 1996).

\bibitem{tegmark93} M.\ Tegmark, ``Apparent wave function collapse caused by scattering''. \textit{Found.Phys.Lett.}\textbf{6,}571(1993).

\bibitem{myatt00} C.\ J.\ Myatt, B.\ E.\ King, Q.\ A.\ Turchette, C.\ A.\ Sackett, D.\ Kielpinski, W.\ M.\ Itano, C.\ Monroe, D.\ J.\ Wineland, ``Decoherence of quantum superpositions through coupling to engineered reservoirs''. \textit{Nature(London).}\textbf{403,}269 (2000).

\bibitem{braginsky87} V.\ B.\ Braginsky, Y.\ I.\ Vorontsov, K.\ S.\ Throne, ``Quantum nondemolition measurenment'', \textit{Science.}\textbf{209,}547(1980).

\bibitem{everett57} H.\ Everett,`` `Relative state formulation' of quantum mechanics", \textit{Rev. Mod. Phys,}\textbf{Vol.29,}454(1957).

\bibitem{monroe96} C.\ Monroe, D.\ M.\ Meekhof, B.\ E.\ King, D.\ J.\ Wineland, ``A ``Schrodinger cat'' superposition state of an atom'', \textit{Science.}\textbf{272,}1131 (1996).

\bibitem{arndi99} M.\ Arndt, O.\ Nairz, J.\ Vos-Andreae, C.\ Keller, G.\ V.\ D.\ Zouw, A.\ Zeilinger, ``Wave-particle duality of C60 molecules''. \textit{Nature(London).}\textbf{401,}680 (1999).

\bibitem{vaidman98} L.\ Vaidman,``On Schizophrenic Experiences of the Neutron or Why We should Believe in the Many-Worlds Interpretation of Quantum Theory", \textit{ International Studies in the Philosophy of Science }\textbf{12,}245 (1998).

\bibitem{tegmark98} M.\ Tegmark,``The Interpretation of Quantum Mechanics: Many Worlds or Many Words?", \textit{Fortschritte der Physik }\textbf{46,}855 (1998).

\bibitem{lockwood96}M.\ Lockwood, `` `Many Mind' interpretation of quantum mechanics'', \textit{Brit.\ J.\ Phil.\ Sci. }, \textbf{47,} 159 (1996).

\bibitem{Wallace02}D.\ Wallace, ``Worlds in the Everett interpretation", \textit{Studies in the History and Philosophy of Modern Physics}\textbf{33,}637 (2002).

\bibitem{Wallace03}D.\ Wallace, `` Everett and Structure ", \textit{Studies in the History and Philosophy of Modern Physics}\textbf{34,}87 (2003).

\bibitem{pribram91} K.\ H.\ Pribram, \textit{Brain and perception, }(Lawrence Erlbaum Associate, 1991).

\bibitem{freeman91} W.\ J.\ Freeman, ``The Physiology of Perception'', \textit{ Sci. Am.}\textbf{264,} 78 (1991).

\bibitem{edelman00} G.\ M.\ Edelman, and G.\ A.\ Tononi, \textit{Universe of Consciousness}(Basic book, 2000).

\bibitem{nicholls00}J.\ G.\ Nichollis, A.\ R.\ Martin, B.\ G.\ Wallace, P.\ A.\ Fuchs, \textit{From Neuron To Brain, fourth edition}(Sinauer Associates, Inc, Publisher, 2000).

\bibitem{llinas85}R.\ R.\ Liinas, ``Electrical Transmission in the Mammalian Central Nervous System'', \textit{ Gap Junction}(Cold Spring Habor Laboratory,1985). 

\bibitem{bennett04}V.\ L.\ Bennett,R.\ S.\ Zukin, ``Electrical coupling and neuronal synchronization in the mammalian brain'', \textit{Neuron}\textbf{41,} 495 (2004).

\bibitem{katz66} B.\ Katz, \textit{Nerve, Muscle, and Synapse}(New York, McGraw-Hill, 1966).

\bibitem{hoppe83} W. \ Hoppe, W.\ Lohmann, H.\ Markl, H.\ Ziegler, \textit{Biophysics}(Springer-Verlog, Berlin, Heidelberg, New York, Tokyo, 1983).

\bibitem{softky93} W.\ R.\ Softky, C.\ Koch, ``The highly irregular firing of cortical cells is inconsistent with temporal integration of random EPSPs'', \textit{Journal of Neuroscience.} \textbf{13,} 334 (1993).

\bibitem{henkel96}A.\ W.\ Henkel,W.\ Almers, ``Faststeps in exocytosis and endocytosis studies by capacitance measurement in endocrine cells'', \textit{Current Opinion in Neurobiology.} \textbf{6,} 350 (1996).

\bibitem{henkel95}A.\ W.\ Henkel, W.\ J.\ Betz, ``Staurosporine blocks evoked release of FM1-43 but not accetylcholine from frog motor nerve terminals'', \textit{ Journal of Neuroscience.} \textbf{15(12),} 8246 (1995).

\bibitem{gandhi03}S.\ P.\ Gandhi, F.\ Stevens, ``Three modes of synaptic vesicular recycling revealed by single-vesicle signaling'', \textit{Nature(London).} \textbf{423,} 607 (2003).

\bibitem{aravanis03}A.\ W.\ Aravanis, J.\ L.\ Pyle, R.\ W.\ Tsien, ``Single synaptic vesicles fusing transiently and successively without loss of identity'', \textit{Nature(London).} \textbf{423,} 643 (2003).

\bibitem{softky95}W.\ R.\ Softky, ``Simple codes versus efficient codes '', \textit{Current Opinion in Neurobiology.} \textbf{5,} 239 (1995).

\bibitem{shadlen94}M.\ N.\ Shadlen, W.\ T.\ Newsome ``Noise, neural codes and cortical organization'', \textit{Current Opinion in Neurobiology.} \textbf{4,} 569 (1994).

\bibitem{zurek00}W.\ H.\ Zurek, ``Einselection and decoherence from an information theory perspective'', \textit{Ann.Phys.} \textbf{9,}853 (2000).

\bibitem{plaga97} R.\ Plaga, ``Proposal for an experimental test of the many-worlds interpretation of quantum mechanics'', \textit{Found.Phys.} \textbf{Vol.27,}559 (1997).

\bibitem{donald90}M.\ J.\ Donald ``Quantum Theory and the Brain'', \textit{Proc.\ R.\ Soc,\ London, Series A.}, \textbf{427}, 43-93 (1990).

\bibitem{beck92} F.\ Beck, J.\ C.\ Eccles, ``Quantum aspects of brain activity and the role of consciousness'', \textit{Proc.\ Nat.\ Acad.\ Sci.\ U.\ S.\ A. }, \textbf{ 89,} 11357 (1992).

\bibitem{roy04} S.\ Roy, M.\ Kafatos, ``Quantum Processes and Functional Geometry: New Perspectives in Brain Dynamics'', \textit{FORMA}, \textbf{19,} 69 (2004).

\end{thebibliography}

\end{document}